\documentclass[a4paper]{PoS}

\usepackage[utf8]{inputenc}
\usepackage{amsmath}
\usepackage{amssymb}
\usepackage{fontenc}
\usepackage{times}
\usepackage{mathptmx}
\usepackage{graphicx}

\newlength{\bibitemsep}
\newlength{\bibparskip}
\let\oldthebibliography\thebibliography
\renewcommand\thebibliography[1]{%
  \oldthebibliography{#1}%
  \setlength{\parskip}{\bibitemsep}%
  \setlength{\itemsep}{\bibparskip}%
}

\title{Static and non-static vector screening masses}

\ShortTitle{Static and non-static vector screening masses}

\author{Bastian~B.~Brandt\\
      \llap{}Institut f\"ur Theoretische Physik, Johann Wolfgang Goethe-Universit\"at,\\Max-von-Laue-Strasse 1, D-60438 Frankfurt am Main, Germany\\
      E-mail: \email{brandt@th.physik.uni-frankfurt.de}}

\author{Anthony~Francis\\
      \llap{}Department of Physics \& Astronomy, York University,\\4700 Keele St., Toronto, ON M3J 1P3, Canada\\
      E-mail: \email{afrancis.heplat@gmail.com}}
      
\author{Harvey~B.~Meyer,~\speaker{Aman~Steinberg}~and~Kai~Zapp\\
      \llap{}PRISMA Cluster of Excellence, Institut f\"ur Kernphysik and Helmholtz-Institut Mainz,\\
	    Johannes Gutenberg-Universit\"at Mainz, D-55099 Mainz, Germany\\
      E-mail: \email{meyerh@uni-mainz.de}, \email{steinber@kph.uni-mainz.de}, \email{zapp@kph.uni-mainz.de}}

\abstract{Thermal screening masses of the conserved vector current are calculated both in a weak-coupling approach and in lattice QCD. The inverse of a screening mass can be understood as the length scale over which an external electric field is screened in a QCD medium. The comparison of screening masses both in the zero and non-zero Matsubara frequency sectors shows good agreement of the perturbative and the lattice results. Moreover, at $T\approx 508\mathrm{MeV}$ the lightest screening mass lies above the free result ($2\pi T$), in agreement with the $\mathcal{O}(g^2)$ weak-coupling prediction.}

\FullConference{34th annual International Symposium on Lattice Field Theory\\
		24-30 July 2016\\
		University of Southampton, UK}

\begin{document}

\section{Introduction}\label{intro}
The thermal screening mass connected to the conserved vector current yields an estimate for the inverse correlation length over which an electric field is screened in a strongly interacting medium, like the quark-gluon plasma (QGP). Screening masses can also be computed perturbatively in the high temperature regime of quantum chromodynamics (QCD) and thus provide a basis for the comparison between lattice and perturbative results. In terms of spectral funtions, the analytical continuation of the screening pole in the Euclidean correlator to the diffusion pole of the retarded correlator establishes a connection of screening masses to real-time quantities, or transport properties, of the QGP \cite{PoSQM2014}. When describing a non-relativistic quark-antiquark pair one encounters the same effective potential that enters the calculation of the dilepton production rate, a real-time quantity of the QGP \cite{screenmassrealtimerates,PoSLat2015}.

For different thermal gauge theories, there are different ways to define the Debye screening mass. In QED, for instance, $k^2+\left. \Pi_{00}(0,k)\right|_{k^2=-m_E^2}=0$ defines the static Debye screening mass $m_E$ as the pole of the longitudinal static photon self-energy \cite{BraatenNieto}. For QCD, however, the chromo-electric Debye mass can be extracted from the correlation function of the imaginary part of the trace of the Polyakov loop which is odd under Euclidean time-reversal \cite{ArnoldYaffe}.

The vector screening mass $M_V$ explored here corresponds to the inverse of the screening length of an external $U(1)$ electric field in the QGP. It can be extracted from the flavour non-singlet vector correlator computed in lattice QCD. The singlet contribution is expected to be small at high temperature.
%

\section{Effective approach}\label{effectiveapproach}
We are interested in investigating the screening of a $U(1)$ electric field in the QGP in terms of the non-singlet vector correlator. The intermediate screening state is described by either an effective field theory or a lattice QCD approach.
In this section, we want to introduce and motivate an effective description of the system, see refs. \cite{screenmassrealtimerates,mesoncorrlengths}. 
The scale hierarchy we can exploit is usually expressed as $g^2T\ll gT\ll 2\pi T$ and separates the non-perturbative \textit{ultrasoft} chromo-magnetic modes at scale $\sim g^2T$ from the intermediate \textit{soft} chromo-electric modes at $\sim gT$ \cite{MikkoBasics}. Both scales are included in the dimensionally-reduced effective theory we will employ, and the \textit{hard} scale $\sim 2\pi T$ enters through perturbative matching. The chromo-electric Debye screening mass $m_E$ associated with the $gT$ scale enters into an effective one-gluon exchange potential. One finds \cite{MikkoBasics}
\begin{eqnarray}
 m_E^2 &=& g^2 T^2 \left(\frac{N_c}{3}+\frac{N_f}{6}\right).
\end{eqnarray}
Following \cite{screenmassrealtimerates}, the thermal flavour non-singlet vector current correlator is defined as
\begin{eqnarray}\label{screencorr}
 G_{\mu\nu}^{(k_n)}(z)&=&\int_0^\beta\mathrm{d}\tau e^{ik_n\tau}\int_{\mathbf{x}}\left\langle\left(\overline{\psi}\gamma_\mu\psi\right)(\tau,\mathbf{x},z)\left(\overline{\psi}\gamma_\nu\psi\right)(0)\right\rangle\,,
\end{eqnarray}
where $\mathbf{x}=(x_1,x_2)^T$ constitutes a transverse plane orthogonal to the $z$ direction. By decomposing the quark fields into their Matsubara modes as
\begin{eqnarray}
 \overline{\psi}(\tau)=T\sum_{p_n}e^{-ip_n\tau}\overline{\psi}_{p_n}\,,\hspace{3mm}\psi(\tau)=T\sum_{p_n}e^{ip_n\tau}\psi_{p_n}\,,
\end{eqnarray}
the screening correlator is re-expressed as
\begin{eqnarray}
 G_{\mu\nu}^{(k_n)}(z)&=&T\int_{\mathbf{x}}\left\langle V_\mu^{(k_n)}(\mathbf{x},z)V_\nu^{(-k_n)}(0)\right\rangle\,,
\end{eqnarray}
where 
\begin{eqnarray}
 V_\mu^{(k_n)}(\mathbf{x},z)&=&T\sum_{p_n}\overline{\psi}_{p_n}(\mathbf{x},z)\gamma_\mu\psi_{p_n-k_n}(\mathbf{x},z).
\end{eqnarray}
\begin{figure}[t]
    \centering
    \includegraphics[width=0.3089\textwidth]{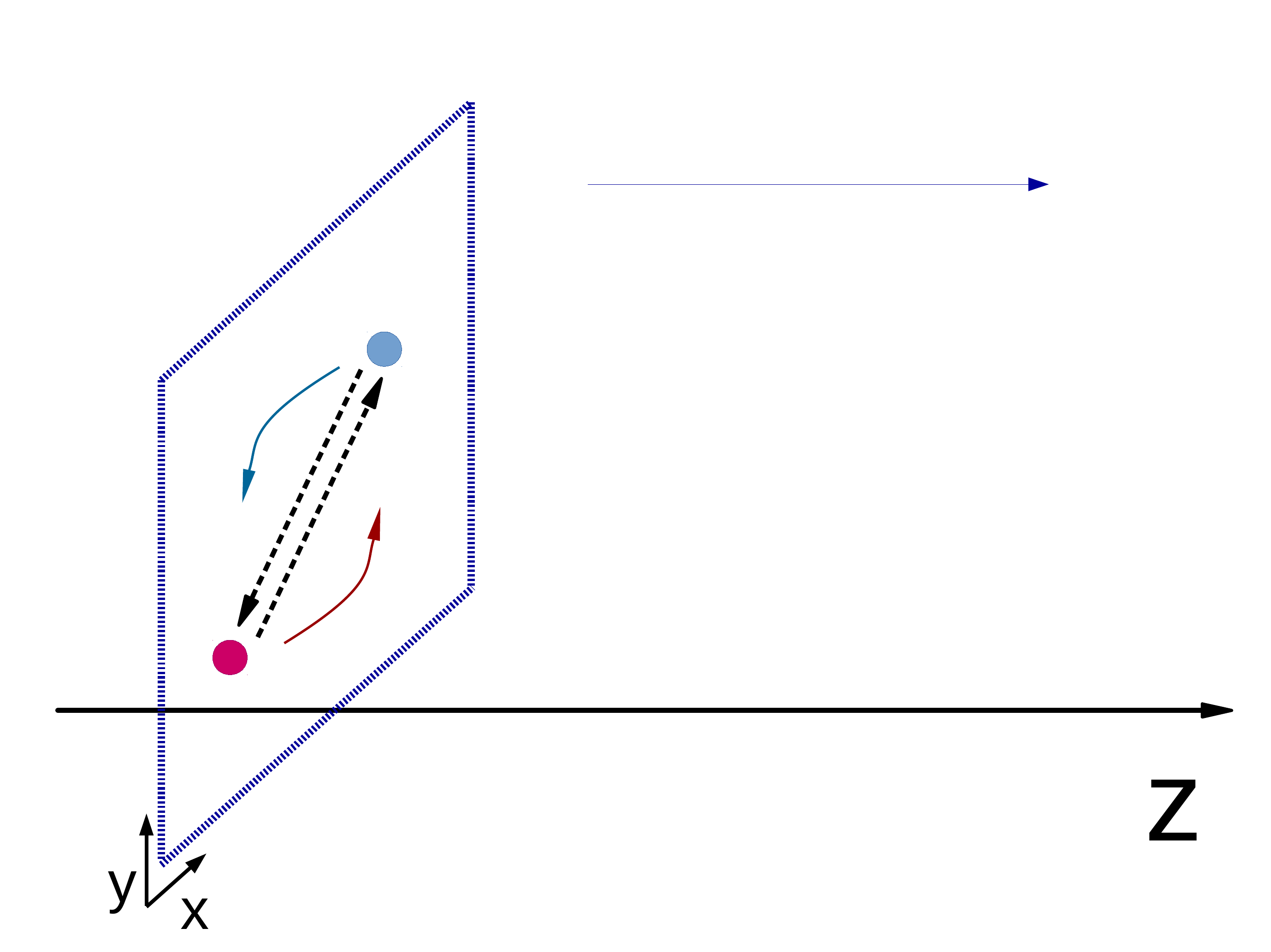}
    \caption{Non-relativistic auxiliary fields in the transverse plane exchanging one gluon. The quark-antiquark pair propagates in the $z$ direction.}
    \label{transverseplaneandamplitudes}
\end{figure}
Fig. \ref{transverseplaneandamplitudes} depicts the quark and antiquark in the transverse plane. With the quark modes carrying frequencies of order $\sim\pi T$, the fields can be dimensionally reduced. As a result we have $2+1$-dimensional non-relativistic fields
\begin{eqnarray}
 \psi=\frac{1}{\sqrt{T}}\begin{pmatrix} \chi \\ \phi \end{pmatrix}.
\end{eqnarray}
This motivates a hydrogen-atom inspired picture for the two-quark bound state in an effective one-gluon exchange potential. The quarks circle each other in the transverse plane orthogonal to the screening direction $z$. Their chromo-electric Debye mass $m_E$ enters in the one-gluon exchange potential as
\begin{eqnarray}
 V_{\mathrm{LO}}^+(\mathbf{y})&=&\frac{g_E^2C_F}{2\pi}\left[\log\left(\frac{m_Ey}{2}\right)+\gamma_E+K_0(m_Ey)\right]\,,
\end{eqnarray}
with $g_E^2=g^2T$ the effective coupling of our dimensionally reduced theory, $C_F=\frac{N_c^2-1}{2N_c}$, $y=\left|\mathbf{y}\right|$ and $K_0$ a modified Bessel function. This potential also enters the calculation of the dilepton production rate as was shown in \cite{screenmassrealtimerates, PoSLat2015} and can be defined non-perturbatively. According to this picture one requires the solution of the radial part of a homogeneous Schr\"odinger equation:
\begin{eqnarray}\label{Seqn}
 \left\lbrace-\frac{d^2}{d\overline{y}^2}-\frac{1}{\overline{y}}\frac{d}{d\overline{y}}+\frac{l^2}{\overline{y}^2}+\rho\left(\frac{2\pi V^{+}_{\mathrm{LO}}}{g_E^2C_F}-\hat{E}^{(l)}\right)\right\rbrace R_l&=&0
\end{eqnarray}
with dimensionless quantities $\overline{y}=m_Ey$, $\rho=g_E^2C_FM_r/(\pi m_E^2)$ and $g_E^2=g^2T$. The first step is to find the (physical) ground-state energy $\hat{E}^{(l)}$ of eq. (\ref{Seqn}), which is used to compute the full energy $E_{\mathrm{full}}$ via
\begin{eqnarray}
 E_{\mathrm{full}}&=&M_{cm}+\frac{g_E^2C_F}{2\pi}\hat{E}^{(l)}\,,\nonumber\\
 M_{cm}&=&k_n+\frac{m_\infty^2}{2M_r}\,,\hspace{3mm} m_\infty^2=\frac{g^2T^2C_F}{4}\,,\hspace{3mm} M_r=\left(\frac{1}{p_n}+\frac{1}{k_n-p_n}\right)^{-1}.
\end{eqnarray}
$E_{\mathrm{full}}$ can now be understood as the screening mass of the $U(1)$ electric field in the medium.

Another interesting object to study is the screening amplitude. The screening correlator exhibits the long-distance behaviour
\begin{eqnarray}
 -\frac{G_{00}^{(k_n)}(z)}{T^3}&\approx&\frac{N_cm_E^2\mathcal{A}_0^+}{\pi T^2}e^{-|z|E_0^{(l=0)}}\,,\nonumber\\
 -\frac{G_{T}^{(k_n)}(z)}{T^3}&\approx&\frac{N_cm_E^4\mathcal{A}_1^+}{\pi T^2}\left[\frac{1}{p_n^2}+\frac{1}{(k_n-p_n)^2}\right]e^{-|z|E_0^{(l=1)}}
\end{eqnarray}
with
\begin{eqnarray}\label{amplitudes}
 \mathcal{A}_0^+=\frac{|R_0(0)|^2}{\int_0^\infty\mathrm{d}\overline{y}\overline{y}|R_0(\overline{y})|^2},\hspace{8mm} \mathcal{A}_1^+=\frac{|R_1'(0)|^2}{\int_0^\infty\mathrm{d}\overline{y}\overline{y}|R_1(\overline{y})|^2}
\end{eqnarray}
for $S$-wave $(l=0)$ and $P$-wave $(l=1)$ channels, respectively.

\begin{figure}[t]
    \centering
    \includegraphics[width=0.18\textwidth]{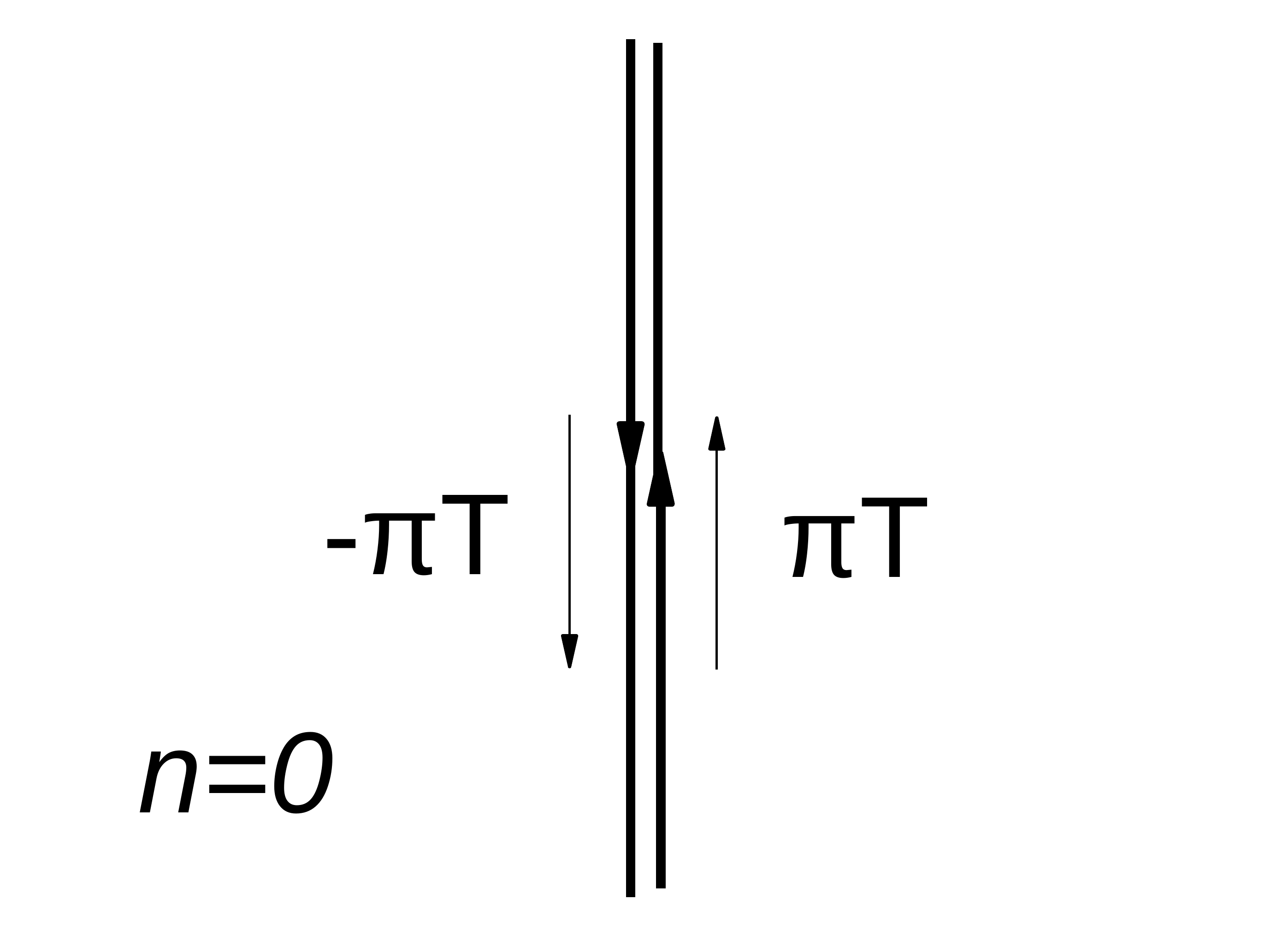}
    \includegraphics[width=0.18\textwidth]{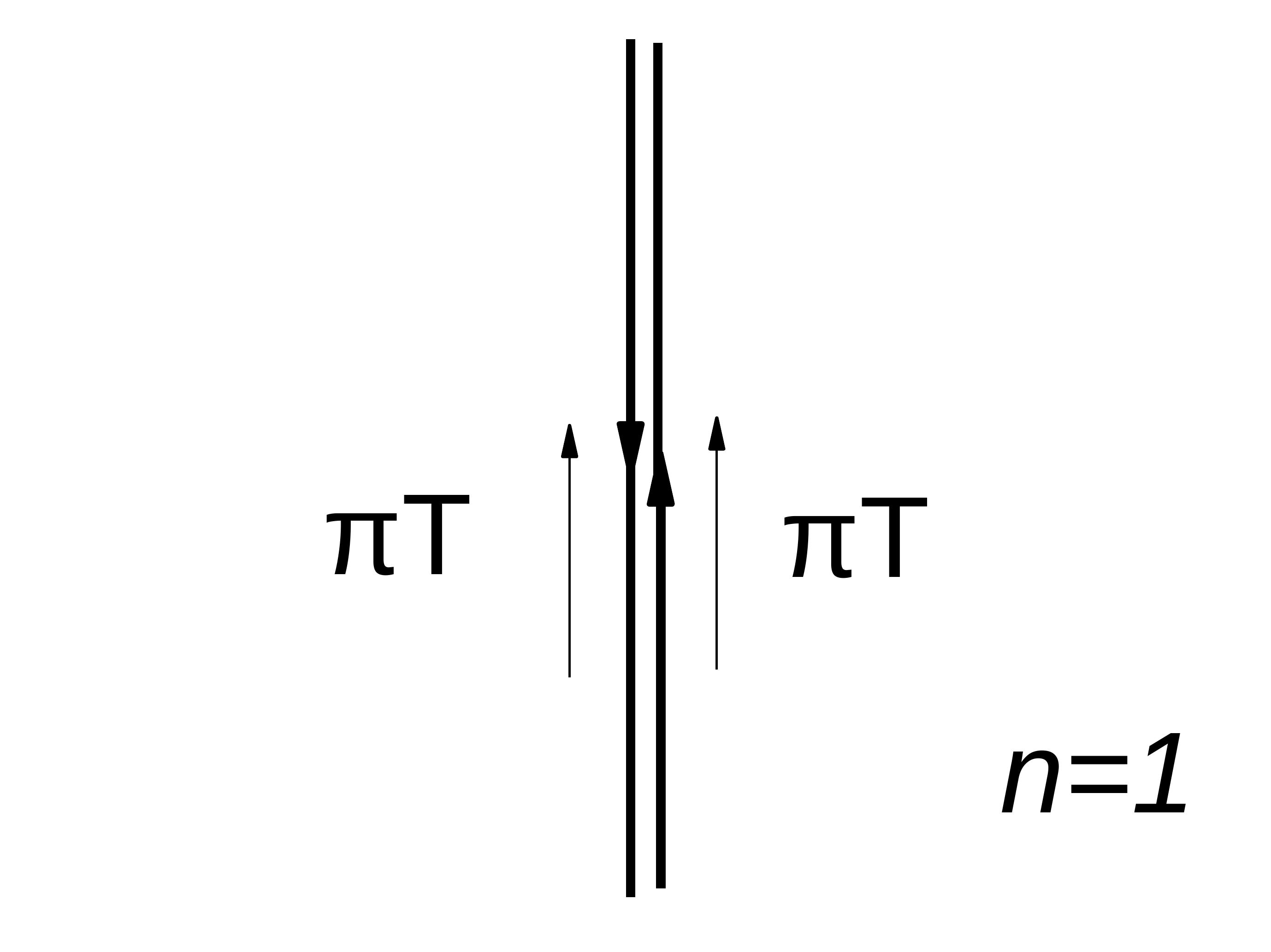}
    \caption{Left: static case ($n=0$) with both quarks carrying momenta of $\pi T$ each in opposing directions. Right: non-static case ($n=1$) with both quarks carrying momenta of $\pi T$ each in the same direction.}
    \label{staticnonstaticquarkpair}
\end{figure}

The situation and the form of the long-distance correlators are quite similar in the static case: both quarks carry momenta of $\pi T$ but in opposite directions as illustrated in fig. \ref{staticnonstaticquarkpair}.
It is important to keep in mind that in the static and the non-static sectors the role of the longitudinal and the transverse parts of the correlator are reversed and it is a different potential that describes the interaction of the quark-antiquark pair.

\section{Lattice calculation}\label{latticesetup}
We simulate the system of interest on a lattice of space-time volume $V=(N_\tau a)\times (N_\sigma a)^3=16a\times (64a)^3$. Using previous lattice simulations with O7 parameters (see \cite{screenmassrealtimerates} and refs. therein) and the running of the coupling $g_0^2$ with the lattice spacing extrapolated from the data in \cite{DellaMorte:2004bc}, the lattice spacing is estimated to be $a\approx 0.024\mathrm{fm}$ corresponding to a temperature of $T=(N_\tau a)^{-1}\cong 508\mathrm{MeV}$. The lattice was generated with $N_f=2$ non-perturbatively $\mathcal{O}(a)$-improved Wilson fermions. We use the plaquette gauge action with $\beta=6/g_0^2=6.038$ \cite{DellaMorte:2004bc}. The critical hopping parameter $\kappa_c$ is extrapolated from the data in \cite{Fritzsch:2012wq} keeping the 2-loop coefficients for $am_c=1/(2\kappa_c)-4$ obtained from \cite{Panagopoulos:2001fn,Luscher:1996vw}. The clover term is set to $c_{\mathrm{sw}}=1.51726$ using the non-perturbative tuning relations of \cite{Jansen:1998mx}. With the chosen hopping parameter $\kappa=0.136238$ we measure an $\overline{\mathrm{MS}}$ mass of $\overline{m}^{\overline{\mathrm{MS}}}(\mu=2\mathrm{GeV})/T\approx0.04$, whereby we follow the conversion of the bare subtracted quark mass to the $\overline{\mathrm{MS}}$ mass of \cite{Fritzsch:2012wq}. The measurements were carried out exploiting $N_{\mathrm{cfg}}=345$ configurations and $N_{\mathrm{src}}=64$ random sources.

We describe the screening correlator of eq. (\ref{screencorr}) by a two-state fit,
\begin{eqnarray}\label{latticefit}
 G_{\mu\nu}^{(k_n)}(z)
 &=&\sum_{n=1}^{2}A_n\frac{\cosh[M_n(z-L_z/2)]}{\sinh[M_n L_z/2]}.
\end{eqnarray}
The effective screening mass $M_1$ we extract from the fit is an estimate for the inverse screening length of the $U(1)$ electric field in the QGP and the excited-state mass $M_2$ corrects for the leading excited-state contamination at long distances.

\begin{figure}[t]
    \centering
    \includegraphics[width=0.49\textwidth]{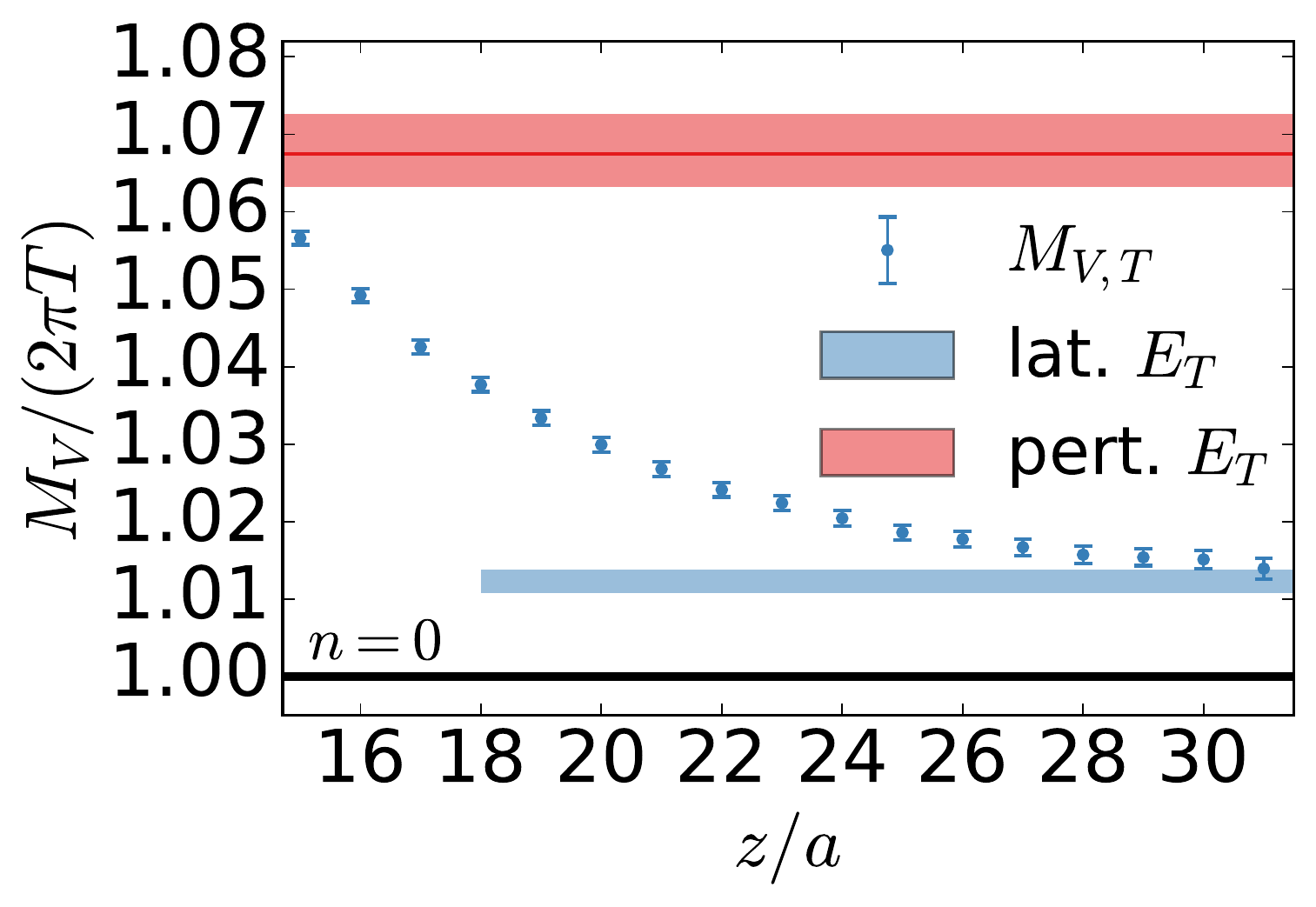}
    \includegraphics[width=0.49\textwidth]{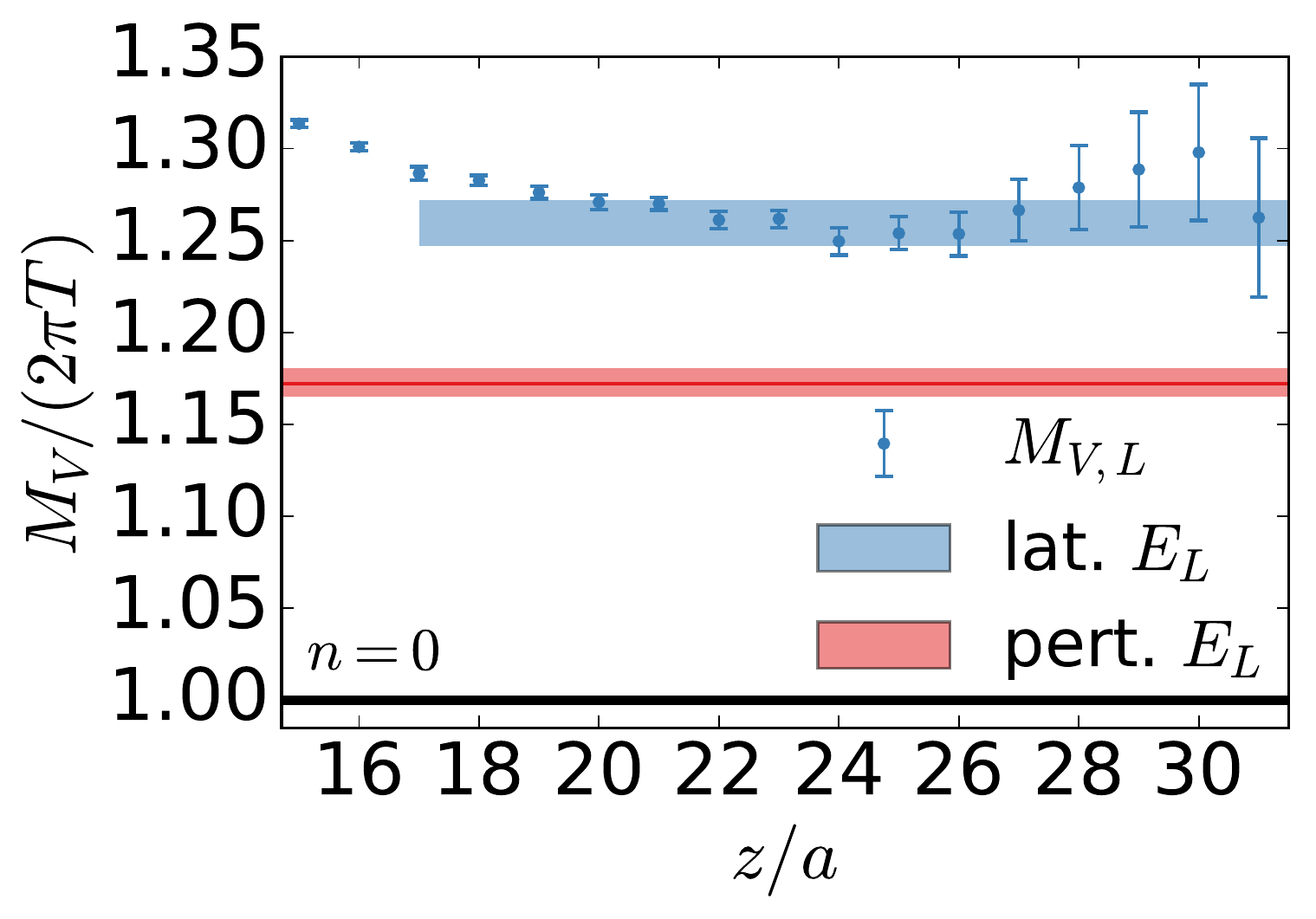}    
    \caption{Static $(n=0)$ screening masses at a temperature of $T=508\mathrm{MeV}$. Left panel: transverse (S-wave) channel. Right panel: longitudinal (P-wave) channel.}
    \label{staticvecscreenmass}
\end{figure}

\begin{figure}[t]
    \centering
    \includegraphics[width=0.49\textwidth]{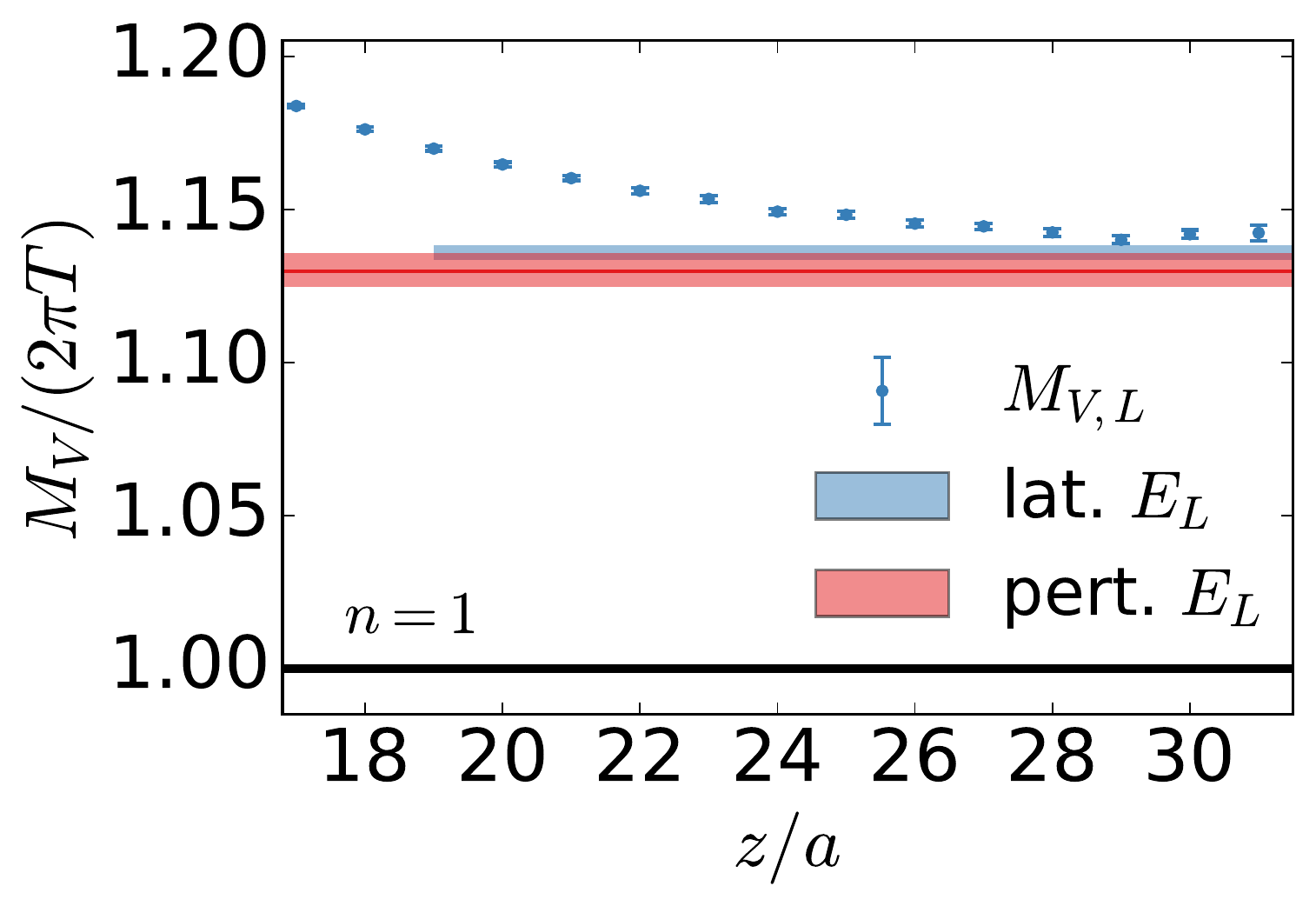}
    \includegraphics[width=0.49\textwidth]{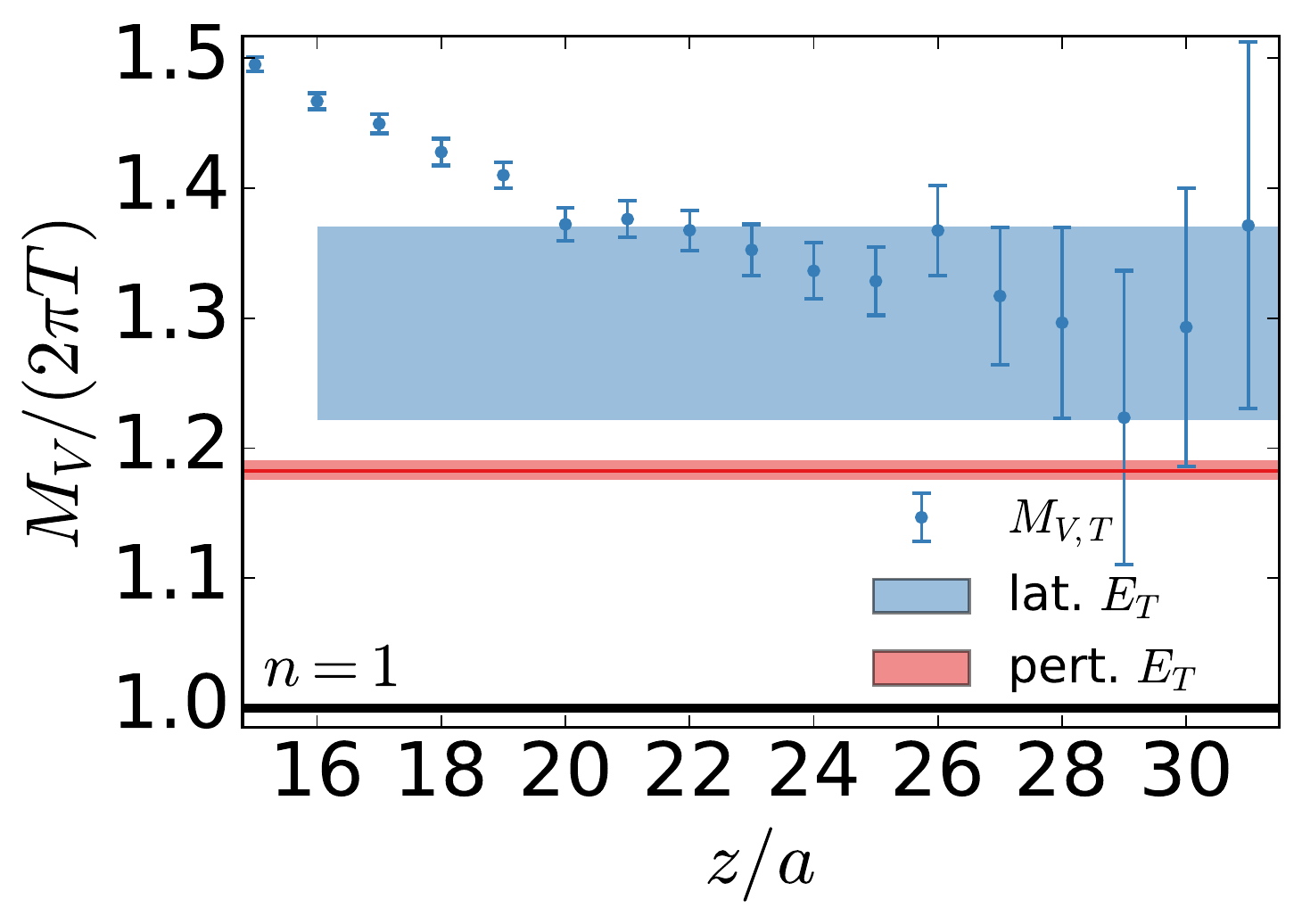}
    \caption{Non-static (n=1) screening masses at a temperature of $T=508\mathrm{MeV}$. Left panel: longitudinal (S-wave) channel. Right panel: transverse (P-wave) channel.}
    \label{nonstaticvecscreenmass}
\end{figure}

We are now in the position to compare the results of the two approaches in order to determine how well the effective perturbative description coincides with the lattice data. In a previous study \cite{screenmassrealtimerates} this comparison was done for temperatures of $T=254\mathrm{MeV}$ and $T=338\mathrm{MeV}$. In the static transverse channel the lattice result lies below the $2\pi T$-line for both temperatures whereas the perturbative result lies above it. The tendency, however, of the lattice data is to eventually cross the $2\pi T$-line at a certain temperature. This behaviour is confirmed by the new lattice data set at $T=508\mathrm{MeV}$, shown on the left of fig. \ref{staticvecscreenmass}. This observation is consistent with the expectation that, the higher the temperature, the more accurate the perturbative description becomes.
The agreement is less good in the static longitudinal channel, see fig. \ref{staticvecscreenmass}, while in both cases the perturbative results agree with the lattice data by less than 10\%. The results obtained by the two different approaches in the non-static longitudinal channel with $n=1$ agree quantitatively, see fig. \ref{nonstaticvecscreenmass} (left). Both values are above $2\pi T$ and are compatible within errors. In the transverse channel of the non-static vector screening correlator the two effective masses are close, although the lattice signal deteriorates as $z$ approaches $L_z/2$, see also fig. \ref{nonstaticvecscreenmass}.
\begin{figure}[t]
    \centering
    \includegraphics[width=0.329\textwidth]{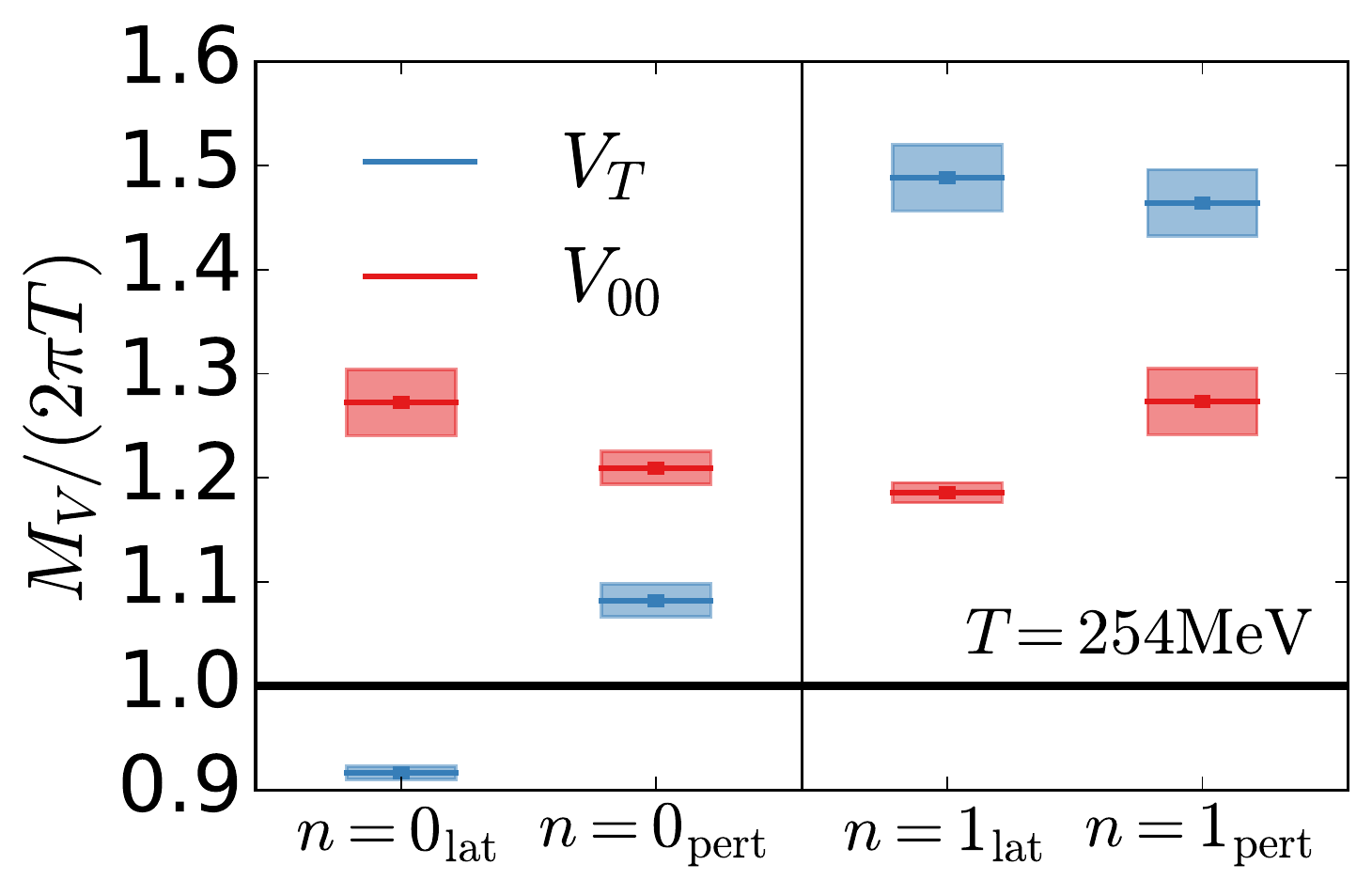}
    \includegraphics[width=0.329\textwidth]{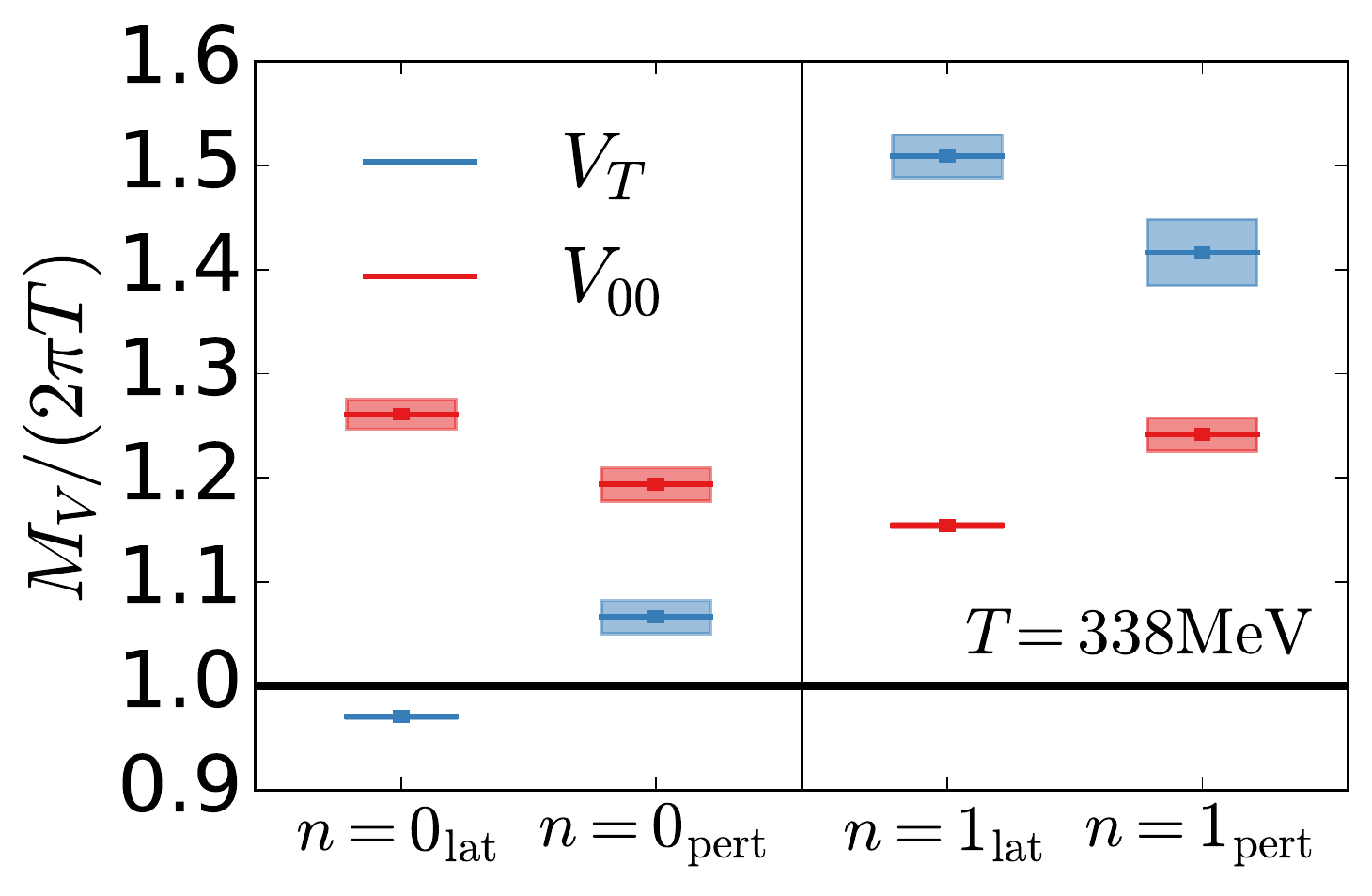}
    \includegraphics[width=0.329\textwidth]{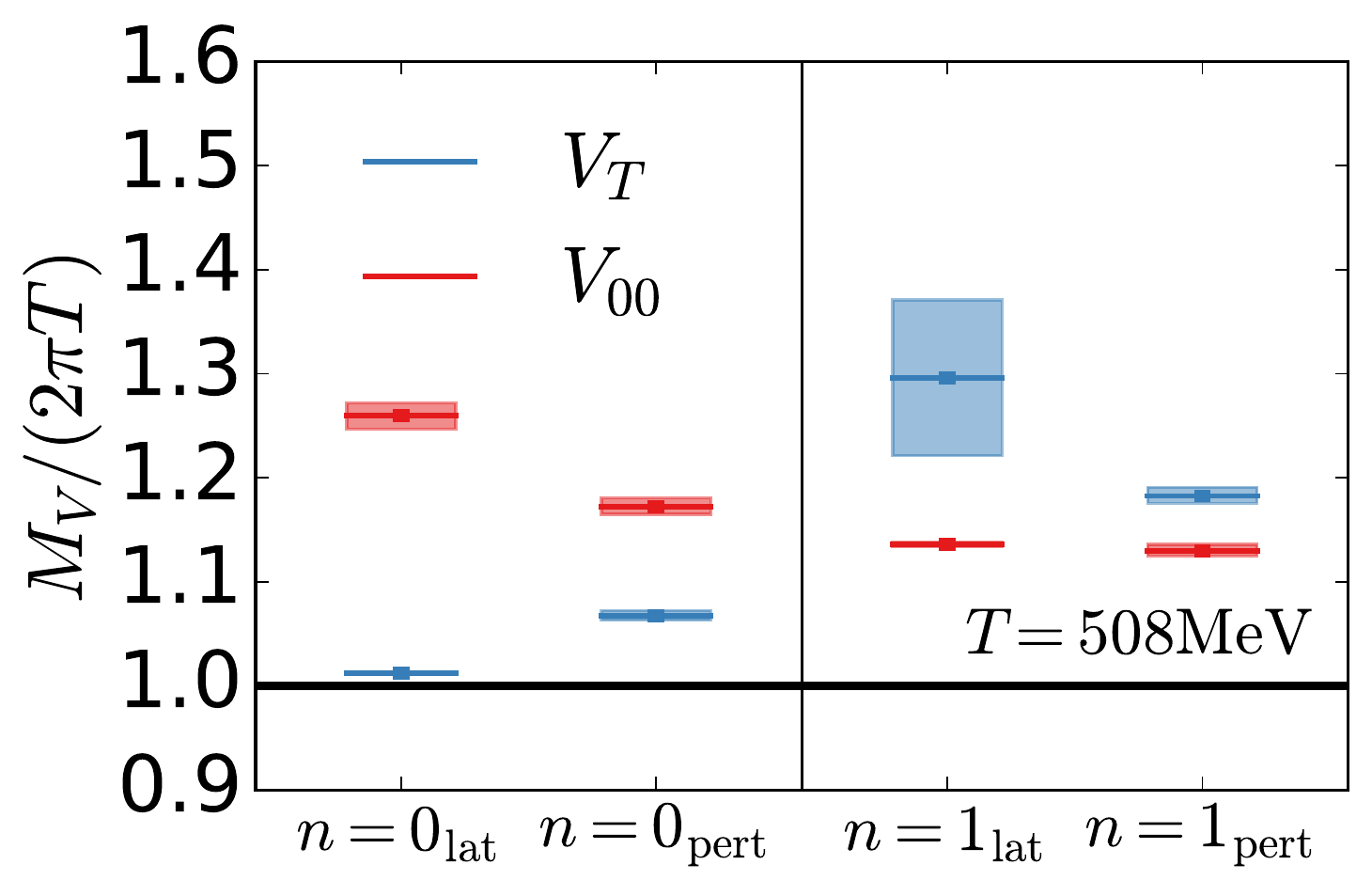}
    \caption{Comparison of the masses at different $T$. Results of the left and middle panels are from \cite{screenmassrealtimerates}.}
    \label{energiescompseveralT}
\end{figure}

Fig. \ref{energiescompseveralT} gives an overview of the spectra at $T=254$, $338$ and $508\mathrm{MeV}$. It is evident that the agreement between lattice field theory and effective field theory is improved for higher temperatures. For the highest temperature, in the static sector, $n=0$, the transverse effective mass lies above the $2\pi T$-line, in qualitative agreement with perturbation theory, and in the non-static sector, $n=1$, the longitudinal effective masses obtained from an effective and a lattice approach agree quantitatively. Since the coupling is smaller for higher temperatures, the difference from $2\pi T$ for the perturbative results decreases with higher temperatures whereas the lattice data stays the same, as can be seen in the longitudinal channel of the static sector, $n=0$. The perturbative results in both the longitudinal and the transverse channel of the non-static sector, $n=1$, at $T=508\mathrm{MeV}$ are systematically shifted to lower values compared to the same data at the lower temperatures $T=254$ and $338\mathrm{MeV}$. For these temperatures a non-perturbative EQCD potential enhanced the agreement of lattice QCD and effective field theory results in the non-static sector, $n=1$, whereas only a leading order potential was available for the comparison at $T=508\mathrm{MeV}$.

\begin{figure}[h]
    \centering
    \includegraphics[width=0.49\textwidth]{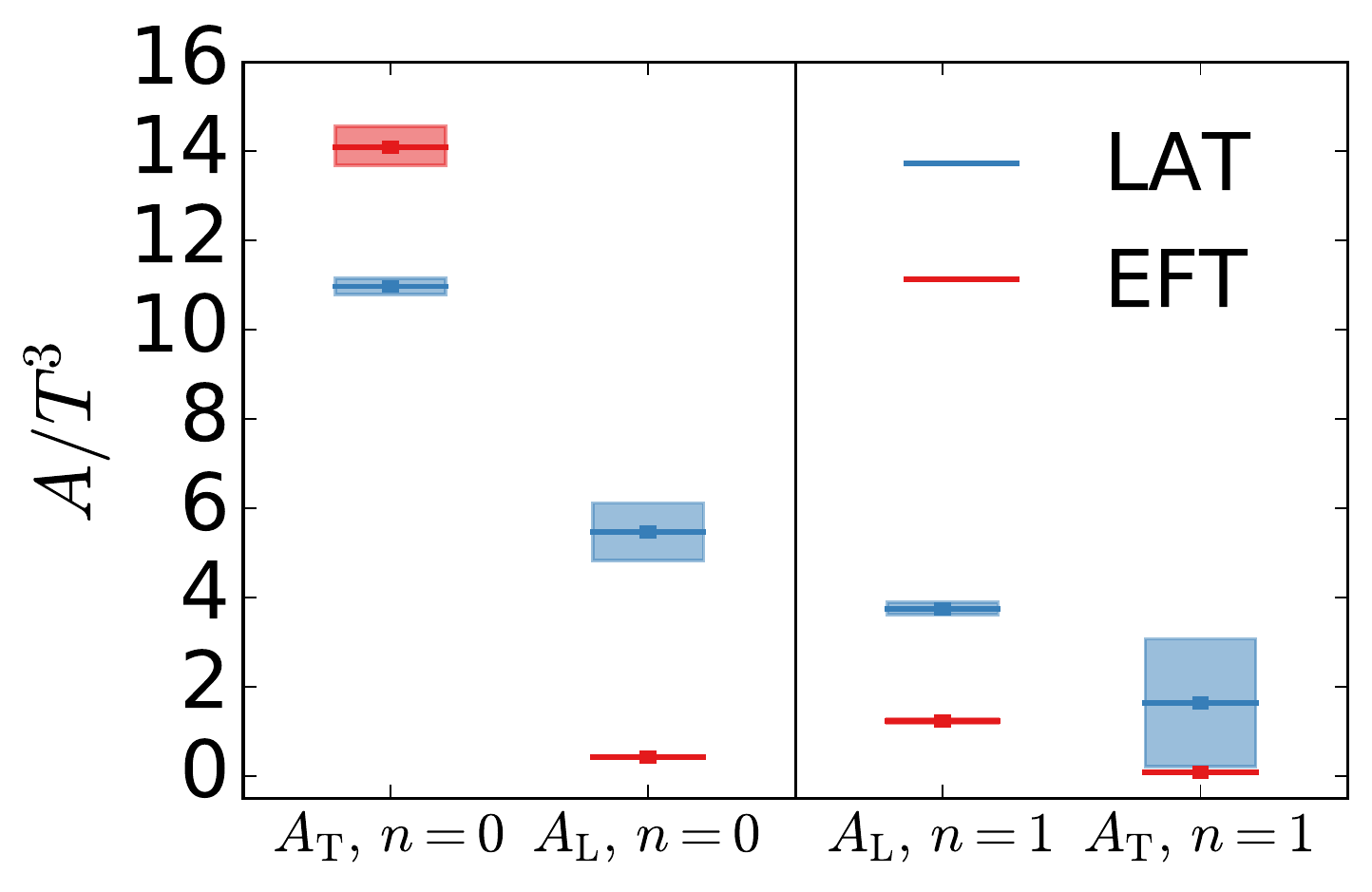}
    \caption{Comparison of the amplitudes at $T=508\mathrm{MeV}$.}
   \label{amplitudescomp508MeV}
\end{figure}

Interpreting the amplitudes, as obtained from eq. (\ref{amplitudes}) and the lattice formulation eq. (\ref{latticefit}), in terms of the quark-antiquark picture, they give an estimate on how tightly bound the quark-antiquark pair is. At the lower temperatures examined in \cite{screenmassrealtimerates} the amplitudes were higher compared to the present study. In this picture, this indicates that at higher temperatures the two quarks are more loosely bound; the bound state is more extended in the $(x,y)$ plane. Although the agreement between lattice and effective field theory is worse than for the masses (see fig. \ref{amplitudescomp508MeV}), there is noticeable improvement when comparing to the results gained for the amplitudes in the previous study \cite{screenmassrealtimerates} at a lower temperature. As compared to the masses, it is more difficult to establish good agreement between lattice and perturbative results for the amplitudes. This is because the leading order value of the mass scales as $2\pi T$ whereas the amplitudes scale as $\sim g^2$ or $\sim g^4$, and are thus more sensitive to uncertainties in the value of the running coupling.

\section{Summary and outlook}\label{sumout}
In this comparative study, both an effective and a lattice approch were implemented in order to examine the effective vector screening mass as the inverse of the effective screening length of a $U(1)$ electric field in the medium. At the temperature of $T=508\mathrm{MeV}$ the lattice result in the transverse channel of the static sector lies above the $2\pi T$-line and therefore agrees better with the perturbative description than at the lower temperatures from a previous study \cite{screenmassrealtimerates}. In the case of the amplitudes of the screening states there is less good agreement between the effective and the lattice theory than for the masses. The agreement is better, however, than for the case of the amplitudes at lower temperatures, see \cite{screenmassrealtimerates}. It was observed earlier, using staggered fermions \cite{screeningmasses2plus1}, that the lowest screening mass lies above the $2\pi T$ threshold for a sufficiently high temperature. The present study confirms this observation on fine lattices and with $\mathcal{O}(a)$-improved Wilson fermions.

A possible future application of screening masses could be their analytic continuation in the Matsubara frequency in order to extract the diffusion coefficient, see \cite{PoSQM2014,PoSLat2015}.

\section{Acknowledgments}\label{ackno}
The authors are thankful for the possibilty to perform the generation of gauge configurations as well as the computation of correlators on the 'Clover' platform at Helmholtz-Institut Mainz. This work was supported in part by DFG Grant ME 3622/2-2. In the course of this work we have also benefitted from resources at FZ Juelich allocated under NIC project HMZ21. The speaker thanks Tim Harris and Daniel Robaina for help and input.


\bibliographystyle{h-physrev5}
\bibliography{pos_lat_2016_steinberg}

\end{document}